\documentclass[preprint]{aastex}
\shorttitle{Six Dwarf Novae}
\shortauthors{Thorstensen et al.}

\begin{document}
\title{Five Dwarf Novae with Orbital Periods Below Two Hours
\footnote{Based on
observations obtained at the Michigan-Dartmouth-MIT Observatory.}
}

\author{John R. Thorstensen and William H. Fenton}
\affil{Department of Physics and Astronomy\\
6127 Wilder Laboratory, Dartmouth College\\
Hanover, NH 03755-3528\\
electronic mail: {\tt john.thorstensen@dartmouth.edu, william.h.fenton@dartmouth.edu}}
\begin{abstract}
We give mean spectra and report orbital periods 
$P_{\rm orb}$ based on
radial velocities taken near minimum light for five 
dwarf novae, all of which prove to have $P_{\rm orb} < 2$ hr.
The stars and their periods are
KX Aql, 0.06035(3) d;
FT Cam, 0.07492(8) d;
% DW Cnc, 0.0599(1) d;
PU CMa, 0.05669(4) d;
V660 Her, 0.07826(8) d;, and
DM Lyr, 0.06546(6).  
The emission lines in KX Aql are notably strong and
broad, and the other stars' spectra appear generally
typical for short-period dwarf novae.   
%The emission velocities in DW Cnc gave unusually
%poor velocity fits despite good signal-to-noise in our
%spectra.  
We observed FT Cam, PU CMa, and DM Lyr on more than
one observing run and constrain their periods accordingly.
%in two runs a month apart, and can constrain its period
%accordingly.  
Differential time-series photometry
of FT Cam shows strong flickering but rules out deep 
eclipses.
%, and comparison with archival data indicates
% a barely significant proper motion around 16 mas yr$^{-1}$.  
Although dwarf novae in this period
range generally show the superhumps and superoutbursts 
characteristic of the SU UMa subclass of dwarf novae,
none of these objects have well-observed superhumps. 

\end{abstract}
\keywords{stars -- individual; stars -- binary;
stars -- variable.}

\section{Introduction}

In this paper we continue determining 
orbital periods $P_{\rm orb}$ for SU-UMa type dwarf novae and candidate
SU UMa stars.
The SU UMa stars are dwarf novae, generally with $P_{\rm orb} < 3$ h,
which occasionally undergo bright and long-duration eruptions,
called superoutbursts.  During superoutburst they show quasi-periodic
photometric oscillations, called superhumps, which 
have periods a few per cent {\it longer} than $P_{\rm orb}$.
\citet{warn} gives an excellent discussion of these stars
(and cataclysmic binaries in general).  At this time the 
most compelling explanation of the superhump clock 
invokes precession of an eccentric accretion disk \citep{whit88}.  
Such disks are expected to develop in high mass ratio (hence short-period)
cataclysmic binaries.

The stars reported on here all prove to have $P_{\rm orb} < 2$ h,
yet to our knowledge none of them have well-observed superhumps
or superoutbursts.  Most of the stars in this sample should show superhumps
in the future.  The superhump periods can then be combined with
the independently determined orbital periods to compute 
the superhump period excess 
$\epsilon = (P_{\rm sh} - P{\rm orb}) / P_{\rm orb}$,
which in turn appears to correlate well with the mass ratio
$q = M_2 / M_1$ \citep{patprecess01}.

\section{Observations}

The time-resolved spectra are from the MDM Observatory
2.4-m Hiltner telescope at Kitt Peak.  Table 1 gives a journal 
of the observations.  We used the modular spectrograph, 
a 600 line mm$^{-1}$ grating, and a $2048^2$ thinned Tektronix
CCD as the detector.  The spectra covered from 4200 to 7500 \AA\ at
almost exactly 2 \AA\ pixel$^{-1}$, with a slightly undersampled
spectral resolution of $\sim 3.5$ \AA\ FWHM.  
Individual exposures were generally $\le 8$ minutes to avoid
phase smearing at short orbital periods, and
frequent comparison exposures maintained the wavelength calibration.  
Observations of flux standard stars yielded a relative flux scale, 
but an unmeasurable and variable fraction of the
light was lost at the $1''$ spectrograph slit, and the
sky was not always photometric, so the
accuracy of the absolute flux calibration is probably 
$\pm \sim 30$ per cent.  The time-averaged, 
flux-calibrated spectra appear in Fig.~1, and 
Table 2 gives measurements of the emission lines.

We used convolution methods \citep{sy} to measure 
velocities of the H$\alpha$ emission lines (Table 3, available
in the electronic edition).  
The velocity uncertainties in 
Table 3 were computed by propagating the counting-statistics estimates 
of the errors in each spectrum channel through the convolution.
%In DW Cnc, H$\alpha$ gave rather poor results despite the
%good signal-to-noise of these spectra, so we also measured
%He I $\lambda$ 5876.
Fig.~2 shows results of period searches on these velocities,
computed using the `residual-gram' method described in  
\citet{tpst}.  
Table 4 gives parameters of sine fits of the form
$$v(t) = \gamma + K \sin[2 \pi (t - T_0) / P]$$
and the RMS scatter $\sigma$ around the best fits.  
In cases where they can be checked, cataclysmic binary 
emission lines seldom reflect the velocity amplitude of the 
white dwarf accurately, so we caution against using 
these parameters to compute masses.  Note that zero phase occurs
at apparent inferior conjunction of the emission-line source.
Fig.~2 also shows the velocities folded on the adopted periods with the
best-fitting sinusoids superposed.

For FT Cam we also obtained differential time-series photometry
with the MDM McGraw-Hill 1.3 m telescope.  A Schott BG38 filter
and a thinned SITe CCD yielded a nonstandard broad blue-visual
passband.  Each image had 256$^2$ binned ($2 \times 2$) pixels,
with an image scale of $1''.018$ per binned pixel.
We obtained 223 exposures of 30 s each, spanning 135 min.
The reduced images were measured using the IRAF\footnote{
The Image Analysis and Reduction Facility is distributed by the National 
Optical Astronomy Observatories.} implementation of 
DAOphot, and the instrumental magnitudes from individual exposures 
were adjusted to correct for transparency variations.  No absolute
calibration was attempted.  An automated routine matched
star images in the pictures to objects in the USNO A2.0 catalog \citep{mon96}
and generated astrometric plate solutions accurate to $\sim 0''.3$.

\section{Notes on Individual Stars}

\subsection{KX Aql}

\citet{tapmenkx} obtained a spectrum of KX Aql showing strong, 
broad emission lines.  The large outburst amplitude
(12.5 to fainter than 18), emission-line morphology, and
absence of a secondary absorption spectrum led them to suggest that
KX Aql is a short period, SU UMa-type system.  \citet{katohvvir} list
it as a candidate WZ Sge-type star.  The period we find,
86.91(3) min, is rather longer than WZ Sge and its ilk.
Also, there is not any clear sign of absorption around 
H$\beta$ from the underlying white dwarf, as is found in 
WZ Sge and GW Lib at minimum light \citep{tpkv02}.

\subsection{FT Cam}

FT Cam was discovered by \citet{antipin99} who designated it 
Var 64 Cam, and found a range from 14.0 to fainter than 17.6 in 
B.  Definite variations were noted around minimum light.  \citet{kato01ft}
reported photometry through an outburst, pointed out that the
known outbursts have all been rare and short in duration,
and derived a proper motion of $\sim 0.02$ arcsec yr$^{-1}$.
Superoutbursts and superhumps have not been detected.

The line spectrum appears typical for a dwarf nova, with a slight
doubling of the emission lines; the peaks of H$\alpha$ are separated
by $\sim 700$ km s$^{-1}$.  In the mean spectrum the violet peaks
of the lines are stronger than the red peaks, which may be an
artifact of uneven phase coverage.  The continuum level implies
$V = 17.5$.   

Our data are from
two observing runs (see Table 1).  The more extensive
2002 January data yield $P_{\rm orb} = 0.07492(8)$ d, but 
the number of cycle counts between this and the 2002 February
run is ambiguous.  Periods within 5 standard deviations of the
best January period are given by $P = [29.621(4)\ {\rm d}] /
N$, where $N = 395 \pm 2$ is an integer.

Figure 3 shows the time-series photometry of FT Cam.
Very strong flickering is present, amounting to almost 1 magnitude
peak-to-peak, but no eclipse is evident.  The flickering is
strong enough that a shallow eclipse might escape
detection.  

Comparing our direct images with the USNO A2.0 yields a barely
significant displacement of FT Cam corresponding to 
$\mu = 16$ mas yr$^{-1}$ in position angle 353$^{\circ}$, 
corroborating the proper motion found by \citet{kato01ft}.

% \subsection{DW Cnc}

% DW Cnc is an unusual and somewhat enigmatic object.  T. Kato
% (2001, VSNET-alert number 5791\footnote{The VSNET home page is
% at {\tt http://vsnet.kusastro.kyoto-u.ac.jp/vsnet/index.html}.  The 
% archived alerts can be found in the ``E-Mail List Archives'' linked
% there.}) informally reported that photometric monitoring shows variations 
% of at least one magnitude, and substantial variability
% on short time scales, though a period could not be
% determined unambiguously from the photometry.   
% 
% DW Cnc is relatively bright -- the 2002 January and February mean spectra 
% indicate $V = 15.5$ and $14.8$ respectively -- so the signal-to-noise is
% quite good.  The Balmer emission lines are almost single-peaked.
% The FeII feature near $\lambda 5169$ is double peaked with
% peaks near $\pm 440$ km s$^{-1}$.  He II $\lambda 4686$ is 
% somewhat more prominent than usual.
% 
% Despite the good signal-to-noise, the radial velocities of 
% H$\alpha$ were somewhat unruly.  After some experimentation we found that
% measurements of the line cores 
% with a convolution function optimized for a 12 \AA\ FWHM
% yielded a usable time series, which gave an unambiguous period
% near 

\subsection{PU CMa}

PU CMa is the optical counterpart of the ROSAT source 
1RXS J064047.8$-$242305.  Several outbursts have been seen, and
in one outburst Kato and Uemura (VSNET alter number 3980) suspected
an oscillation which might have been the beginnings of a superhump.
\footnote{The VSNET home page is
{\tt http://vsnet.kusastro.kyoto-u.ac.jp/vsnet/index.html}}
.  

The emission lines are only slightly double-peaked in the 
average spectrum, indicating a rather low orbital inclination.
The observed continuum implies $V = 16.2$, quite bright for a
relatively unstudied object.  The radial velocity data span
a 6.8-h range in hour angle, despite the unfavorable declination
of $-24^{\circ}$; this and the good signal-to-noise resulted
in an unambiguous period determination.  The period, 
81.63(6) min, is the shortest in the present sample and 
is essentially equal to that of WZ Sge.  The low velocity
amplitude $K$ (Table 3) also suggests a relatively low
inclination.

As with FT Cam, the data set spans two 
observing runs, creating ambiguity
in the long-term cycle count.  The more extensive 2002 January data
yield $P = 0.05669(4)$ d.  Including the 2002 February data
constrains the period to $[29.5941(32)\ \rm{d}] / N$,
where the integer $N =  522 \pm 1$.

\subsection{V660 Her}

\citet{spogli98} obtained photometry during an outburst which
reached $V = 14.3$; at minimum the magnitude is {\it estimated} as
19, so the amplitude is approximately 5 mag. 
A spectrum obtained by \citet{liu99} appears normal for a dwarf nova
near minimum light.  Our spectrum (Fig.~1) appears similar, with
single-peaked Balmer lines which suggest a relatively
low orbital inclination.  The continuum indicates $V \sim 18.7$.
The period, 112.60(12) min, is the longest in the present sample.

\subsection{DM Lyr}

The emission lines are single-peaked, indicative of a low
orbital inclination.  
DM Lyr is the only star in this sample for which we are 
aware of a superhump period, namely 0.066 d listed by 
\citet{nogami97} in their Table 1.  However, the details 
of this determination remain
unpublished, and the quoted accuracy is insufficient 
to compute an accurate superhump period excess $\epsilon$.  
Our spectroscopy is from 
two observing runs separated by two years, which yielded
a weighted mean $P_{\rm orb} = 0.06546(6)$ d.  Precise
periods which fit both runs lying within four standard 
deviations of the weighted mean are given by 
$[747.9533(25)\ {\rm d}] / N$, where the integer 
$N = 11426 \pm 44$.

\section{Discussion}

None of these five stars appears particularly unusual.
They all prove to have periods in the range
occupied by the SU UMa stars.
Superhumps are evidently detected (but not
well measured) in DM Lyr, and presumably they have
not yet turned up in the other four stars
only because the objects have not been observed long or intensively
enough.  If one or more of the other
four objects proves after extensive monitoring {\it not} to be 
an SU UMa star, it will present an interesting anomaly.
%As noted earlier, the superhump period excess $\epsilon = 
%(P_{\rm sh} - P_{\rm orb}) / P_{\rm orb}$ appears to give a 
%measure of the mass ratio $q = M_2 / M_1$ \citep{pattprecess01}.

{\it Acknowledgments.} We thank the NSF for support through 
AST 9987334.   Tim Miller obtained the direct images of FT Cam.
This research made use of the Simbad database,
operated at CDS, Strasbourg, France.

\clearpage
\begin{deluxetable}{lrrrlrrr}
\tabletypesize{\scriptsize}
\tablewidth{0pt}
\tablecolumns{8}
\tablecaption{Journal of Spectroscopy}
\tablehead{
\colhead{Date} & \colhead{N} & \colhead{HA start} & \colhead{HA end} &
\colhead{Date} & \colhead{N} & \colhead{HA start} & \colhead{HA end} \\ 
% \colhead{Date} & \colhead{N} & \colhead{HA start} & \colhead{HA end} \\
\colhead{(UT)} & \colhead{} &\colhead{(hh:mm)} & \colhead{(hh:mm)} & 
% \colhead{(UT)} & \colhead{} &\colhead{(hh:mm)} & \colhead{(hh:mm)} & 
\colhead{(UT)} & \colhead{} &\colhead{(hh:mm)} & \colhead{(hh:mm)} 
}
\startdata
 {\it KX Aql:}&  &  &  &                     {\it V660 Her:}  \\
 2001 Jun 24 &  3 & $ +1:21$ & $ +1:35$ &   2001 Jun 24 &  4 & $ -0:41$ & $ +2:56$  \\  
 2001 Jun 26 &  2 & $ +0:35$ & $ +0:46$ &   2001 Jun 25 &  5 & $ -1:50$ & $ -1:12$  \\  
 2001 Jun 30 & 20 & $ -3:07$ & $ +2:36$ &   2001 Jun 26 & 20 & $ -1:33$ & $ +4:13$  \\ 
 2001 Jul  1 & 26 & $ -4:11$ & $ +2:52$ &   2001 Jun 27 & 15 & $ -2:21$ & $ -0:23$  \\ 
 {\it FT Cam:} &  &  &  &                   2001 Jun 28 &  5 & $ +2:04$ & $ +2:39$  \\ 
 2002 Jan 22 & 31 & $ +0:11$ & $ +5:54$ &   2001 Jun 29 &  6 & $ -0:20$ & $ +0:23$  \\ 
 2002 Jan 23 & 34 & $ -0:51$ & $ +5:58$ &   2001 Jul  1 &  3 & $ +3:52$ & $ +4:10$  \\ 
 2002 Jan 24 &  8 & $ +0:14$ & $ +1:00$ &   {\it DM Lyr:} \\
 2002 Feb 20 &  2 & $ +1:46$ & $ +1:52$ &   1999 Jun 4 &  3 & $ -0:38$ & $ +0:27$ \\  
 2002 Feb 21 &  4 & $ +2:22$ & $ +4:19$ &   1999 Jun 9 &  9 & $ +0:59$ & $ +1:57$ \\  
 {\it PU CMa:}& &  &  &                     1999 Jun 10 & 19 & $ -4:49$ & $ +2:11$ \\ 
 2002 Jan 20 & 17 & $ -2:47$ & $ +2:45$  &  1999 Jun 11 &  6 & $ -0:25$ & $ +0:14$ \\ 
 2002 Jan 21 & 22 & $ -3:07$ & $ +2:48$  &  1999 Jun 12 & 12 & $ +0:46$ & $ +2:17$ \\ 
 2002 Jan 22 &  4 & $ +1:19$ & $ +1:36$  &  1999 Jun 13 & 12 & $ -2:44$ & $ -0:56$ \\ 
 2002 Jan 23 & 18 & $ -3:11$ & $ +3:38$  &  1999 Jun 14 &  6 & $ -1:17$ & $ -0:33$ \\ 
 2002 Feb 19 &  2 & $ +0:28$ & $ +0:38$  &  2001 Jun 27 & 16 & $ -0:59$ & $ +2:27$ \\ 
 2002 Feb 20 &  4 & $ -0:25$ & $ +1:43$  &  2001 Jun 28 &  5 & $ -3:39$ & $ -3:04$ \\ 
\enddata
\end{deluxetable}
\clearpage

\begin{deluxetable}{lrcc}
\tablewidth{0pt}
\tablecolumns{4}
\tablecaption{Emission Features}
\tablehead{
\colhead{Feature} & 
\colhead{E.W.\tablenotemark{a}} & 
\colhead{Flux\tablenotemark{b}}  & 
\colhead{FWHM \tablenotemark{c}} \\
 & 
\colhead{(\AA )} & 
\colhead{(10$^{-15}$ erg cm$^{-2}$ s$^{1}$)} &
\colhead{(\AA)} \\
}
\startdata
\cutinhead{KX Aql:} 
           H$\gamma$ & $104$ & $225$ & 27 \\ 
            H$\beta$ & $128$ & $213$ & 26 \\ 
  HeI $\lambda 4921$ & $ 10$ & $ 15$ & 28 \\ 
  HeI $\lambda 5015$ & $ 14$ & $ 22$ & 28 \\ 
   Fe $\lambda 5169$ & $ 12$ & $ 19$ & 29 \\ 
  HeI $\lambda 5876$ & $ 52$ & $ 58$ & 32 \\ 
           H$\alpha$ & $227$ & $225$ & 30 \\ 
  HeI $\lambda 6678$ & $ 23$ & $ 23$ & 37 \\ 
  HeI $\lambda 7067$ & $ 14$ & $ 15$ & 34 \\ 
\cutinhead{FT Cam:}
           H$\gamma$ & $ 54$ & $376$ & 29 \\ 
  HeI $\lambda 4471$ & $ 15$ & $ 91$ & 35 \\ 
  HeII$\lambda 4686$ & $  7$ & $ 39$ & 46 \\ 
            H$\beta$ & $ 85$ & $418$ & 31 \\ 
  HeI $\lambda 4921$ & $ 10$ & $ 47$ & 41 \\ 
  HeI $\lambda 5015$ & $ 11$ & $ 53$ & 38 \\ 
   Fe $\lambda 5169$ & $  9$ & $ 42$ & 35 \\ 
  HeI $\lambda 5876$ & $ 37$ & $132$ & 38 \\ 
           H$\alpha$ & $145$ & $483$ & 36 \\ 
  HeI $\lambda 6678$ & $ 17$ & $ 53$ & 41 \\ 
  HeI $\lambda 7067$ & $ 14$ & $ 41$ & 45 \\ 
% \cutinhead{DW Cnc (2002 January):}
%           H$\gamma$ & $ 65$ & $2588$ & 24 \\ 
%  HeI $\lambda 4471$ & $ 16$ & $568$ & 21 \\ 
%  HeII $\lambda 4686$ & $  9$ & $307$ & 23 \\ 
%            H$\beta$ & $ 84$ & $2608$ & 25 \\ 
%  HeI $\lambda 4921$ & $  7$ & $216$ & 23 \\ 
%  HeI $\lambda 5015$ & $  8$ & $221$ & 24 \\ 
%   Fe $\lambda 5169$ & $  3$ & $ 95$ & 26 \\ 
%  HeI $\lambda 5876$ & $ 22$ & $540$ & 26 \\ 
%           H$\alpha$ & $ 94$ & $2043$ & 26 \\ 
%  HeI $\lambda 6678$ & $ 11$ & $227$ & 28 \\ 
%  HeI $\lambda 7067$ & $  8$ & $157$ & 30 \\ 
\cutinhead{PU CMa:}
           H$\gamma$ & $ 68$ & $1637$ & 23 \\ 
  HeI $\lambda 4471$ & $ 20$ & $386$ & 26 \\ 
  HeII $\lambda 4686$ & $ 11$ & $198$ & 44 \\ 
            H$\beta$ & $108$ & $1784$ & 25 \\ 
  HeI $\lambda 4921$ & $  9$ & $147$ & 30 \\ 
  HeI $\lambda 5015$ & $  9$ & $131$ & 27 \\ 
   Fe $\lambda 5169$ & $  4$ & $ 59$ & 27 \\ 
  HeI $\lambda 5876$ & $ 38$ & $455$ & 30 \\ 
           H$\alpha$ & $151$ & $1674$ & 27 \\ 
  HeI $\lambda 6678$ & $ 19$ & $201$ & 34 \\ 
  HeI $\lambda 7067$ & $ 14$ & $139$ & 36 \\ 
\cutinhead{V660 Her:}
           H$\gamma$ & $ 52$ & $125$ & 26 \\ 
  HeI $\lambda 4471$ & $ 14$ & $ 27$ & 24 \\ 
  HeII $\lambda 4686$ & $ 15$ & $ 24$ & 55 \\ 
            H$\beta$ & $ 88$ & $133$ & 26 \\ 
  HeI $\lambda 4921$ & $  9$ & $ 13$ & 31 \\ 
  HeI $\lambda 5015$ & $ 10$ & $ 15$ & 27 \\ 
   Fe $\lambda 5169$ & $  6$ & $  8$ & 30 \\ 
  HeI $\lambda 5876$ & $ 24$ & $ 29$ & 28 \\ 
           H$\alpha$ & $107$ & $122$ & 27 \\ 
  HeI $\lambda 6678$ & $ 12$ & $ 14$ & 35 \\ 
  HeI $\lambda 7067$ & $  9$ & $ 10$ & 42 \\ 
\cutinhead{DM Lyr:}
           H$\gamma$ &  86: & 61: & 15 \\
  HeI $\lambda 4471$ &  15: & 14: & 21 \\
            H$\beta$ & $ 78$ & $ 79$ & 18 \\
  HeI $\lambda 4921$ & $  8$ & $  8$ & 18 \\
  HeI $\lambda 5015$ & $  8$ & $  8$ & 19 \\
   Fe $\lambda 5169$ & $  8$ & $  8$ & 18 \\
  HeI $\lambda 5876$ & $ 27$ & $ 24$ & 23 \\
           H$\alpha$ & $ 95$ & $ 82$ & 20 \\
  HeI $\lambda 6678$ & $ 13$ & $ 11$ & 23 \\
  HeI $\lambda 7067$ & $  9$ & $  7$ & 25 \\
\enddata
\tablenotetext{a}{Emission equivalent widths are counted as positive.}
\tablenotetext{b}{Absolute line fluxes are uncertain by a factor of about
2, but relative fluxes of strong lines
are estimated accurate to $\sim 10$ per cent.} 
\tablenotetext{c}{From Gaussian fits.}
\end{deluxetable}

\clearpage
\begin{deluxetable}{lrr}
\tablewidth{0pt}
\tablecolumns{3}
\tablecaption{H$\alpha$ Radial Velocities}
\tablehead{
\colhead{Time} & \colhead{$v$} & \colhead{$\sigma$\tablenotemark{a}} \\ 
\colhead{(HJD 24$\ldots$)} & \colhead{km s$^{-1}$} & \colhead{km s$^{-1}$} 
}
\startdata
\cutinhead{KX Aql:}
52084.92859 &  $-66$ &  9  \\
52084.93335 &  $-39$ &  9  \\
52084.93810 &  $ -4$ &  8  \\
52086.89135 &  $ 19$ & 20  \\
52086.89883 &  $ 16$ & 25  \\
52090.72698 &  $-20$ & 10  \\
\enddata
\tablenotetext{a}{Standard deviation estimated from counting statistics.}
\end{deluxetable}

\begin{deluxetable}{lllrrcrccc}
\tabletypesize{\footnotesize}
\tablewidth{0pt}
\tablecaption{Fits to Radial Velocities}
\tablehead{
\colhead{Data set} & \colhead{$T_0$\tablenotemark{a}} & \colhead{$P$} &
\colhead{$K$} & \colhead{$\gamma$} & \colhead{$\sigma$}  & \colhead{$N$} & 
Funct.\tablenotemark{b} &
$W$ & $F$ \\
\colhead{} & \colhead{} &\colhead{(d)} & \colhead{(km s$^{-1}$)} &
\colhead{(km s$^{-1}$)} & \colhead{(km s$^{-1}$)} & & & \colhead{(\AA )} &
\colhead{(\AA )}
}
\startdata
KX Aql & 52090.7923(9) & 0.06035(2) &  52(5) & $-11(4)$ & 51 &  19 & G & 30 & 12\\ 
FT Cam & 52298.6667(11) & 0.074990(10)\tablenotemark{c} &  56(5) & $ 6(4)$ & 79 & 18 
& G & 46 & 12\\
% DW Cnc: H$\alpha$ & 52321.7017(13) & 0.059933(6)\tablenotemark{b} &  117(14) & $ 34(10)$ & 124 &  60 \\
% DW Cnc: HeI & 52321.7118(9) & 0.059794(3)\tablenotemark{b} &  179(17) & $ 47(12)$ & 127 &  69 \\
% PU CMa & 52295.8772(8) & 0.05670(4) &  36(3) & $-24(2)$ & 61 &  13 & G & 44 & 12\\ 
PU CMa & 52297.9183(9) & 0.056694(6)\tablenotemark{c} &  37(4) & $-25(3)$ & 67 & 14 & 
G & 44 & 12\\ 
V660 Her & 52087.6838(17) & 0.07826(8) &  39(5) & $ 35(4)$ & 58 &  17 & D & 36 & \nodata\\
DM Lyr & 51343.8565(15) & 0.0654092(2)\tablenotemark{c} &  37(5) & $-36(4)$ & 88 &  21
& G & 30 & 10 \\ 
\enddata
\tablenotetext{a}{Blue-to-red crossing, HJD $- 2400000$.}
\tablenotetext{b}{Convolution function used in measurements; $D$ indicates the
derivative of a gaussian, in which case $W$ is the FWHM for which it is 
optimized, while $G$ indicates positive and negative gaussians of FWHM $F$ separated
by $W$.}
\tablenotetext{c}{For these objects, the period and its error assumes an
uncertain choice of cycle count between observing runs. See text for discussion.}
\end{deluxetable}

\clearpage

\begin{figure}
\plotone{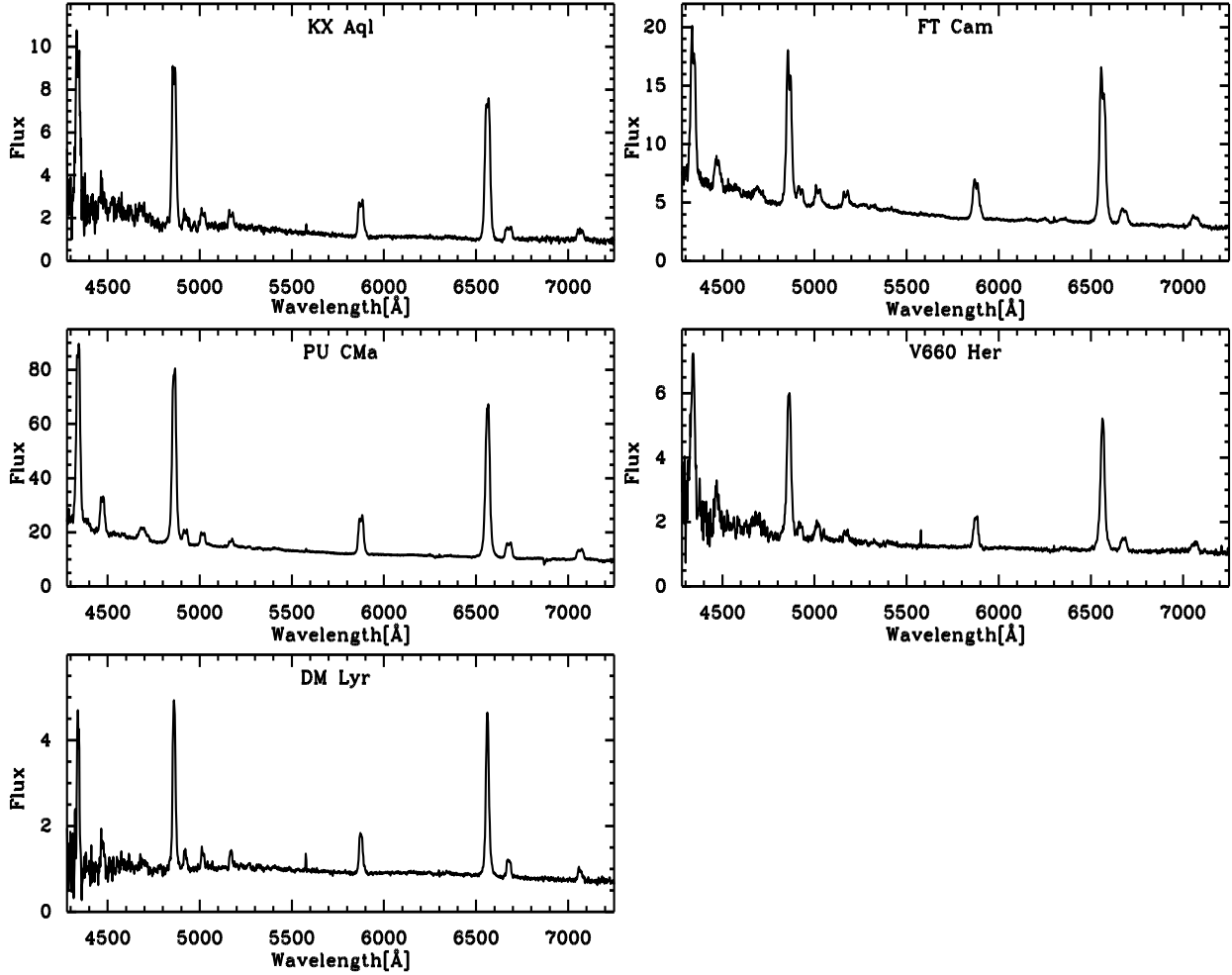}
\caption{Averaged spectra.  The vertical axes
are in units of $10^{-16}$ erg cm$^{-2}$ s$^{-1}$ \AA$^{-1}$, but
the flux scales are uncertain by at least 20 per cent.
}
\end{figure}
\clearpage

\begin{figure}
\epsscale{0.83}
\plotone{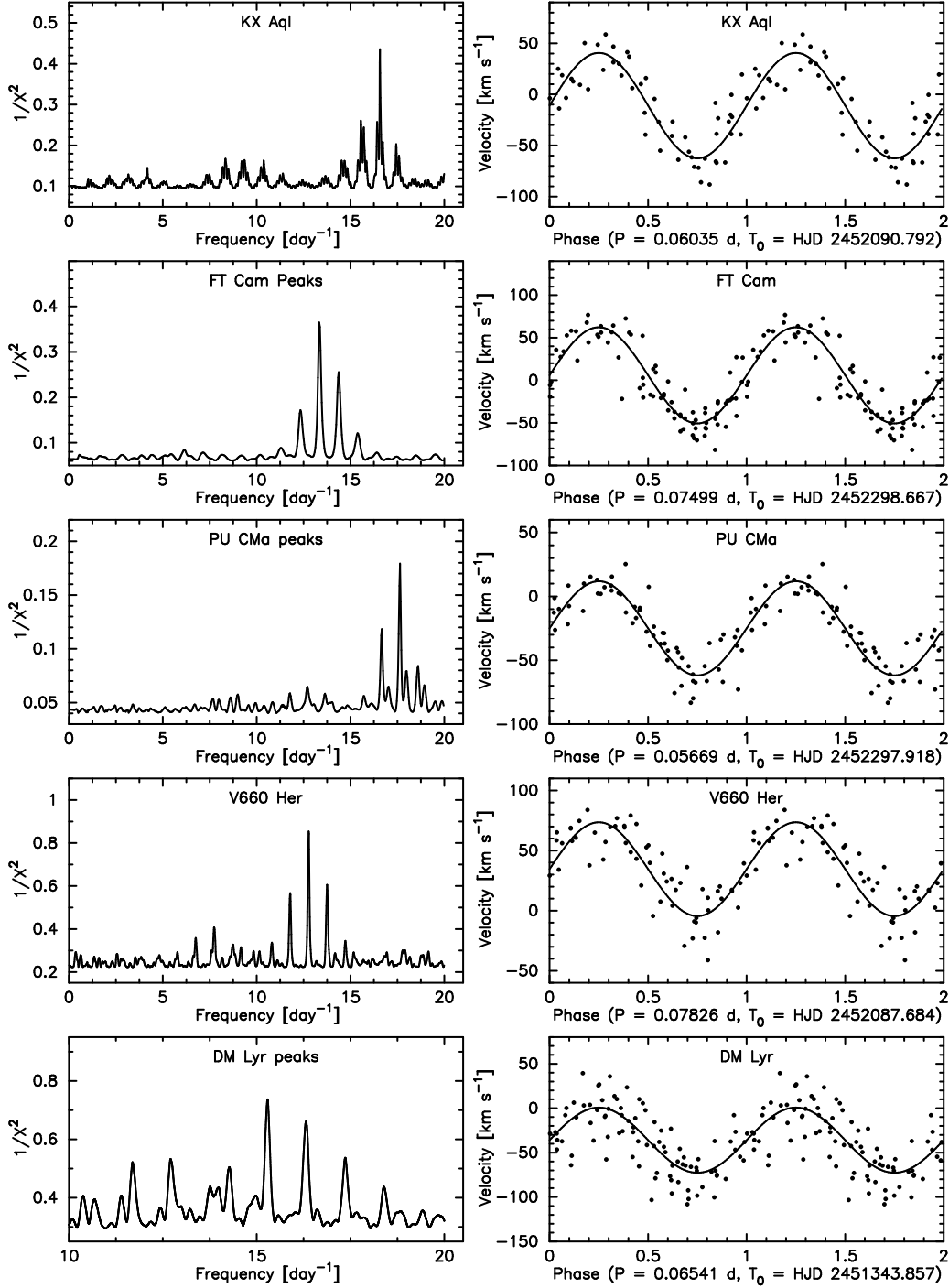}
\caption{Period searches of the radial velocities 
(left panels), and radial velocities folded on the adopted 
periods (right panels).  When
data from several observing runs are combined, fine-scale
ringing is present, and the function plotted is formed by 
joining local maxima of the function.  In these cases the 
choice of cycle count used in folding the data for the right
panel is arbitrary.  Two cycles are shown in the folds for
continuity.
}
\end{figure}
\clearpage

% \begin{figure}
% \figurenum{2}
% \plotone{fig2b.ps}
% \caption{ (Continued).
% }
% \end{figure}
% \clearpage

\begin{figure}
\plotone{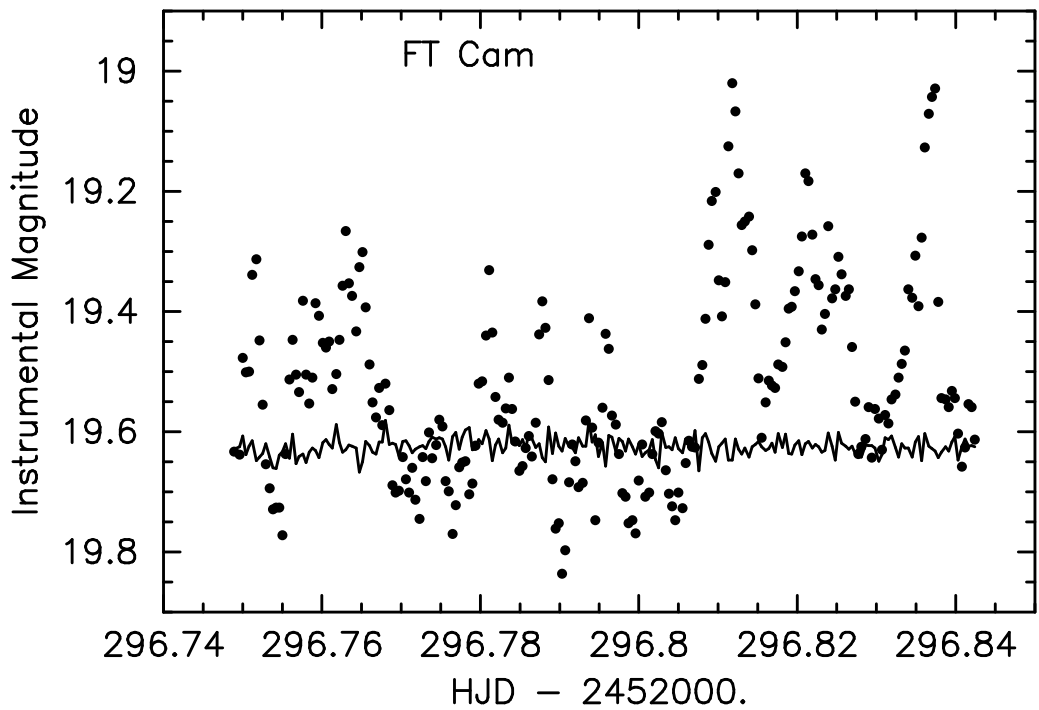}
\caption{Time-series photometry of FT Cam.  Instrumental
magnitudes from each frame have been adjusted to account
for transparency variations.  The jagged line shows 
a comparison star at $\alpha = 3^{\rm h}\ 20^{\rm m}\ 
58^{\rm s}.88, \delta = 61^{\circ}\ 05'\ 13''.8$ (ICRS),
that is, 113 arcsec from FT Cam in PA 264 degrees.
}
\end{figure}
\clearpage

%\begin{figure}
% \plotone{fig4.ps}
%\caption{The Stolz-Schoembs relation; see the discussion
%in the text.  The dashed line is an empirical fit to a
%subset of the data.
%}
%\end{figure}
%\clearpage


\clearpage
\begin{thebibliography}

\bibitem[Antipin(1999)]{antipin99} Antipin, S. V. 1999, Inf. Bull. Var. Stars, 
No. 4673

\bibitem[Kato, Uemura, \& Yamaoka(2001)]{kato01ft} Kato, T., Uemura, M., \&
Yamaoka, H. 2001, Inf. Bull. Var. Stars, No. 5082

\bibitem[Kato, Sekine, \& Hirata(2001)]{katohvvir} Kato, T., 
Sekine, Y., \& Hirata, R. 2001, \pasj, 53, 1191

\bibitem[Liu et al.(1999)]{liu99} Liu, W., Hu, J. Y., Zhu, X. H., 
and Li, Z. Y. 1999, \apjs, 122, 243

\bibitem[Monet et al.(1996)]{mon96} Monet, D. et al. 1996,
USNO SA-2.0, (U. S. Naval Observatory, Washington, DC)

% sx lmi Psh, mentions DM Lyr low-accuracy Psh in a table.
% no longer cited if DM Lyr is out.
\bibitem[Nogami, Masuda, \& Kato(1997)]{nogami97} Nogami,
D., Masuda, S., and Kato, T. 1997, PASP, 109, 1114

\bibitem[Patterson(2001)]{patprecess01} Patterson, J. 2001, \pasp, 113, 736

\bibitem[Schneider \& Young(1980)]{sy} Schneider, D. and Young, P. 
1980, ApJ, 238, 946

\bibitem[Spogli, Fiorucci, \& Tosti(1998)]{spogli98} 
Spogli, M. Fiorucci, M., and Tosti, G. 1998, A\&AS, 130, 485

\bibitem[Tappert \& Mennickent(2001)]{tapmenkx} 
Tappert, C., and Mennickent, R. E. 2001, Inf. Bull. Var. Stars,
No. 5101

\bibitem[Thorstensen et al.(1996)]{tpst} Thorstensen, J. R., Patterson, J., Shambrook, A., and
Thomas, G.  1996, PASP 108, 73

\bibitem[Thorstensen et al.(2002)]{tpkv02} Thorstensen, J. R., Patterson, J. O., Kemp, J.,
and Vennes, S. 2002, \pasp, 118, in press

\bibitem[Warner(1995)]{warn} Warner, B. 1995, Cataclysmic Variables (Cambridge University
Press)

\bibitem[Whitehurst(1988)]{whit88} Whitehurst, R. 1988, \mnras, 232, 35


\end{thebibliography}
\end{document}